\documentclass[amssymb,amsmath,preprint]{revtex4-1}

\usepackage{graphicx}
\usepackage{epsfig}

\newcommand{\nsub}[1]{_{\rm{#1}}}

\begin{document}


\title{The Cold Neutron Chopper Spectrometer at the Spallation Neutron Source 
-- A Review of the first 8 Years of Operation}

\author{G. Ehlers}
\affiliation{Quantum Condensed Matter Division, Oak Ridge National
Laboratory, Oak Ridge, TN 37831, USA}
\author{A. A. Podlesnyak}
\affiliation{Quantum Condensed Matter Division, Oak Ridge National
Laboratory, Oak Ridge, TN 37831, USA}
\author{A. I. Kolesnikov}
\affiliation{Chemical and Engineering Materials Division, Oak Ridge 
National Laboratory, Oak Ridge, TN 37831, USA}


\begin{abstract}
The first eight years of operation of the Cold Neutron Chopper Spectrometer 
(CNCS) at the Spallation Neutron Source in Oak Ridge is being reviewed. 
The instrument has been part of the facility user program since 2009, 
and more than 250 individual user experiments have been performed to date. 
CNCS is an extremely powerful and versatile instrument and offers leading 
edge performance in terms of beam intensity, energy resolution, 
and flexibility to trade one for another. 
Experiments are being routinely performed with the sample at extreme 
conditions: $T\lesssim{0.05}$~K, $p\gtrsim{2}$~GPa and $B=8$~T can be achieved 
individually or in combination.
In particular, CNCS is in a position to advance the state of the art with 
inelastic neutron scattering under pressure, and some of the recent 
accomplishments in this area will be presented in more detail.

\vspace{1in}

\footnotesize{Note: This manuscript has been authored by UT-Battelle, LLC 
under Contract No. DE-AC05-00OR22725 with the U.S. Department of Energy.  
The United States Government retains and the publisher, by accepting the 
article for publication, acknowledges that the United States Government retains 
a non-exclusive, paid-up, irrevocable, world-wide license to publish or 
reproduce the published form of this manuscript, or allow others to do so, 
for United States Government purposes.  
The Department of Energy will provide public access to these results of 
federally sponsored research in accordance with the DOE Public Access Plan 
(http://energy.gov/downloads/doe-public-access-plan).}

\end{abstract}

\maketitle


\section{Introduction}
\label{Intro}

The Cold Neutron Chopper Spectrometer (CNCS) is one of the direct geometry, 
inelastic neutron scattering spectrometers that the Spallation Neutron Source 
(SNS) in Oak Ridge operates as part of its facility user program. 
This instrument has now been operating for about 8 years. 
It has seen a number of incremental upgrades during this time, which have 
resulted in an overall significant improvement of its performance. 
In this paper the most important of these developments are being described, 
and some of the science is being reviewed that has been done with the 
instrument. 

\begin{figure*}[t]
\begin{center}
\includegraphics[width=6in]{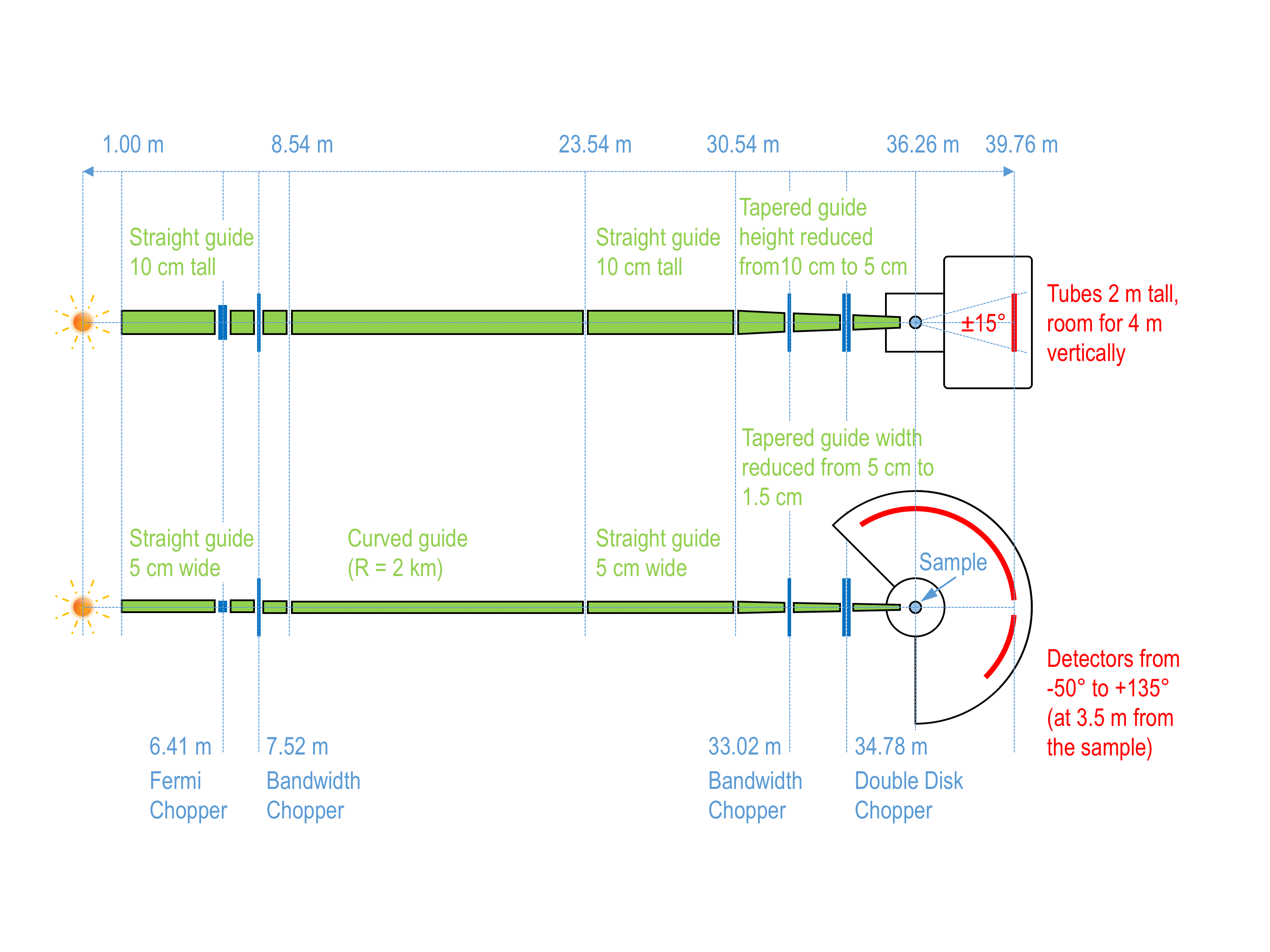}
\end{center}
\caption{Schematic layout of CNCS, viewed from the side (top part of figure) 
and from the top (bottom part of figure). 
The location of various key components is indicated relative to the 
moderator surface. 
The drawing is not to scale.}
\label{schematic}
\end{figure*}

CNCS is a general-purpose direct-geometry inelastic time-of-flight (TOF) 
spectrometer optimized for cold neutrons.~\cite{Ehl11}  
It complements the other direct geometry time-of-flight instruments 
at the SNS: The wide angular range chopper spectrometer (ARCS), 
the fine-resolution Fermi chopper spectrometer (SEQUOIA), 
and the hybrid spectrometer (HYSPEC).~\cite{Sto14}  
The general layout of the instrument is shown in Fig.~\ref{schematic}.
CNCS receives beam from a cold coupled moderator (liquid H$_2$) with a peak 
brightness at $\sim{10}$~meV neutron energy.~\cite{Ive03} 
Short pulses of a monochromatic neutron beam are directed onto the sample, 
and the location (scattering angle) and time-of-flight of the detected 
neutrons are used to determine the energy and the momentum exchanged 
between neutron and sample in the scattering event.
A pair of high speed choppers is used to select the neutron energy $E\nsub{i}$ 
via their relative phase: a Fermi chopper at 6.41~m (distance from the 
moderator surface) and a high-speed double-disk (HSDD) chopper at 34.78~m. 
The energy resolution is mostly determined by the double disk chopper 
settings, and can be varied independently of the chosen value of $E\nsub{i}$. 
Each of the two HSDD chopper disks has three slits with different 
widths~\cite{Ehl11}, giving the operator many options to change the burst 
time by re-phasing to a different slit, or by changing the chopper speed.
Two bandwidth choppers at 7.52~m and at 33.02~m cut out unwanted neutrons 
from other frames. 
The neutron guide is mostly straight, but its central part is horizontally 
curved to bring the sample position out of direct line-of-sight from the source. 
Thus the guide acts as a filter for high energy ($\gtrsim{100}$~meV) neutrons 
which cannot make the required reflections in the supermirror coating. 

Neutron data are recorded in event mode, which is now becoming the standard 
method at pulsed spallation neutron sources. 
For each detected neutron, the time, pulse ID and pixel ID are stored in a 
list. 
The pulse ID refers to the source pulse that generated the neutron, and has to 
be determined carefully when the instrument (like CNCS) does not always 
operate in the first frame (which is to say, when not all neutrons have 
been detected by the time the next pulse hits). 
Complementing slow controls data (such as temperatures, motor positions, 
chopper phases, etc.) are all stored with time stamps which enables one 
to correlate them with counted neutrons during or after acquisition. 
Taking data in event mode is a very powerful approach, as it offers a maximum 
of flexibility to filter and analyze the data after the acquisition ended. 
In event mode it is also straightforward to perform pump-probe and 
time-resolved experiments.
High performance software packages exist to reduce and scientifically 
analyze the data.~\cite{dave,mantid,horace}


\section{Modifications since start of Operation}
\label{Mods}

\subsection{Focussing guide end section}

\begin{figure*}[t]
\begin{center}
\includegraphics[width=3.5in]{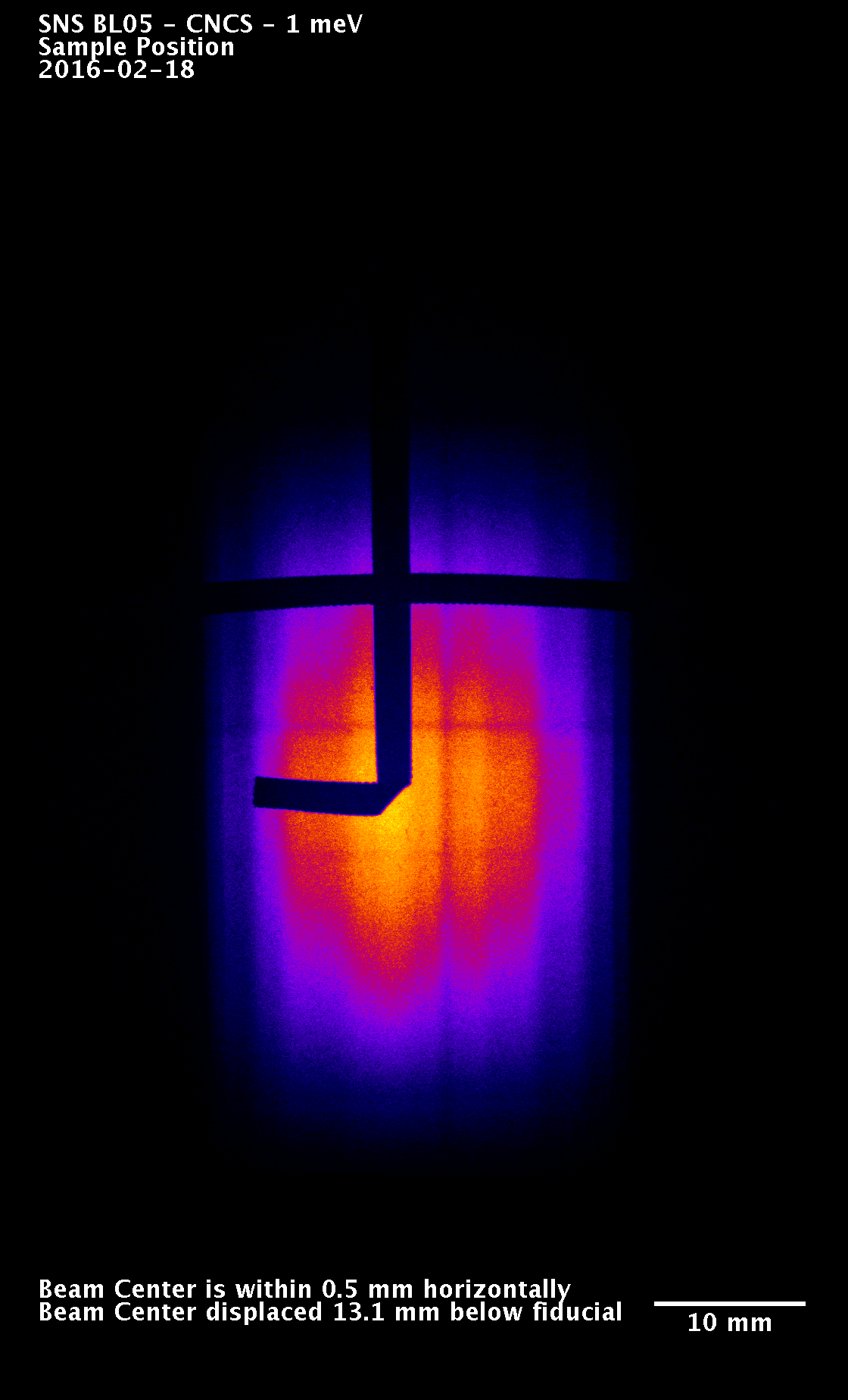}
\end{center}
\caption{A high resolution photograph of the beam at the sample position, 
taken with the new focussing guide end piece. 
Comparing with the beam from the traditional, not focussing guide end section 
(see Fig. 2 in Ref.~\cite{Ehl11}), this beam is considerably more compressed 
in the vertical direction. 
The fiducial was placed for reference only. 
This picture was taken with 1~meV neutrons.}
\label{beam}
\end{figure*}

A new guide end section has been available since 2014. 
This device makes use of $m=6$ supermirrors, which were not commercially 
available at the time CNCS was designed and built. 
It optionally replaces the previously existing, $\sim{20}$~cm long last guide 
section directly in front of the sample position.
Due to their parabolic shape the mirrors focus the beam at the sample position 
to a $\sim{20\times{15}}$~mm spot (height $\times$ width). 
A high resolution image of the beam at the sample position, with the new guide 
section, is shown in Fig.~\ref{beam}. 
This is a fairly small beam for this type of instrument, but it corresponds 
well to the typical sample dimensions which are $\lesssim{1}$~cm for single 
crystals (see section~\ref{use} below). 
Compared to the traditionally used last guide section~\cite{Ehl11}, the 
corresponding intensity gain for small crystal samples is about a factor of 
$\sim{4}$, which comes mostly at the cost of increased vertical divergence. 
Since the vertical direction (in which the $Q$ resolution is broadened) lines 
up well with Debye Scherrer cones, the new guide is also very beneficial for 
powder samples, even when they are large enough to be not fully illuminated by 
the beam.
This proves to be a crucial intensity gain, in particular in experiments 
with small samples which are the vast majority (see section~\ref{use} below).
The new guide end section does impact the horizontal divergence only 
marginally, which is set by the $m=2-3.5$ mirrors further upstream in the 
guide. 
The `new' and `old' guide end pieces can be readily exchanged. 
While the `new' guide is also most often used for polycrystalline and liquid 
samples, the `old' guide is preferred in experiments that aim to measure 
excitations in single crystal samples with good $Q$ resolution in all three 
spatial directions simultaneously. 
Such experiments account for $\sim{10}-20$\% of beam time use. 
Most experiments with crystals, however, focus on the scattering in one plane, 
which is arranged to be horizontal in the laboratory frame.  
In this situation the vertical $Q$ resolution is of lesser importance and the focusing guide section is thus preferred.


\subsection{Fermi chopper rotor}

The original CNCS design placed a high value on achieving very good energy 
resolution. 
This can be seen, for example, in the guide design. 
The guide narrows down significantly towards its end, which allows one to 
achieve very short burst times with the double disk chopper. 
As a result, the beam at the sample position is quite narrow, $\sim{15}$~mm, 
which limits the size of single-crystal samples one can measure (crystals are 
being rotated during a measurement which implies that they should be fully 
illuminated by the beam because an intensity change upon rotation is hard 
to correct for).
Secondly, in order to achieve very good resolution, the Fermi chopper in the 
front was originally equipped with a rather tight slit package. 

\begin{figure*}[t]
\begin{center}
\includegraphics[width=6in]{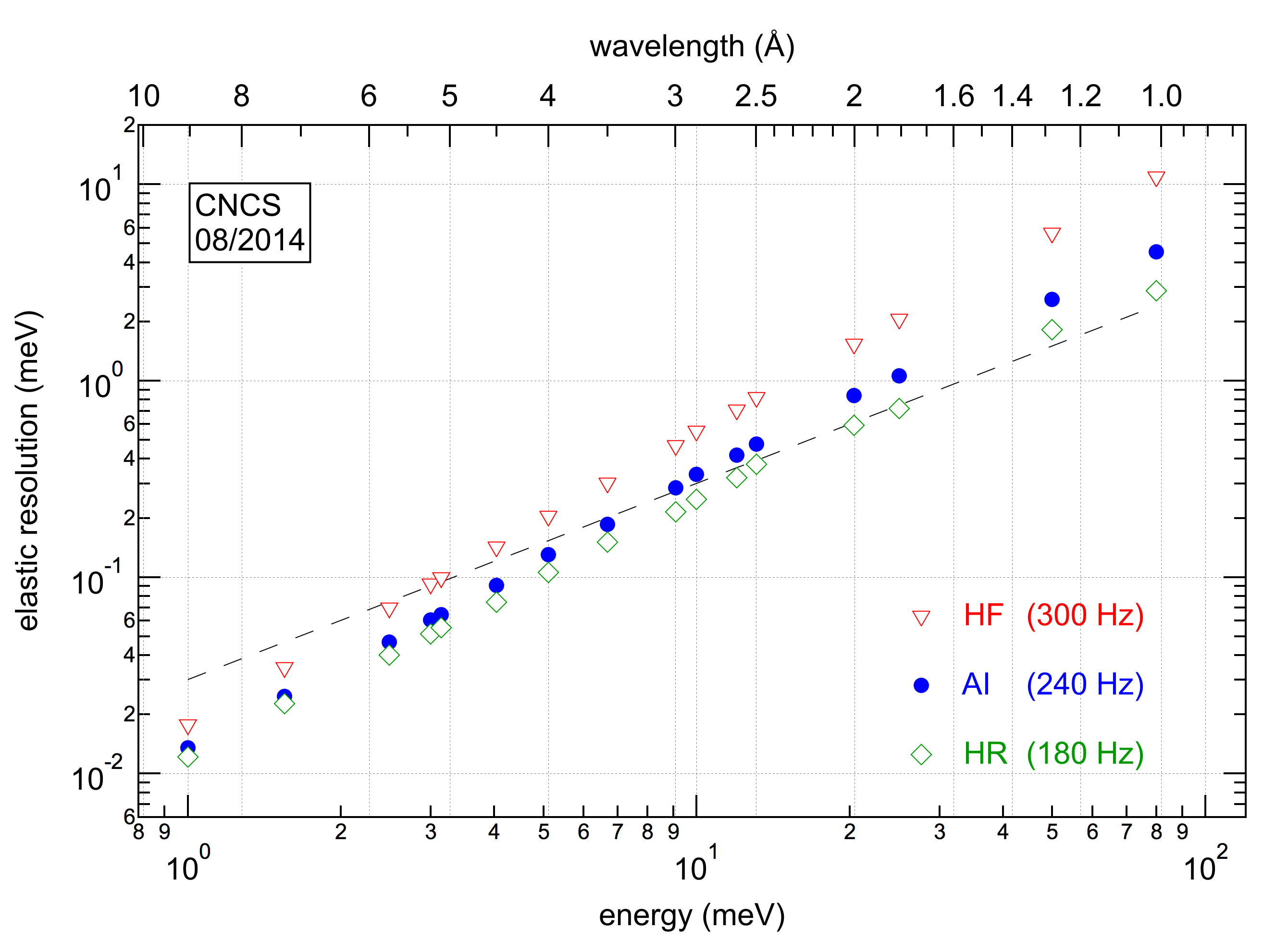}
\end{center}
\caption{A current measurement (with the new Fermi chopper rotor) of the 
energy resolution at the elastic line, using a vanadium reference sample. 
Three representative settings have been chosen which cover most of the 
available range of resolution settings. 
A 2.5\% resolution is indicated by the dashed line. 
Going from HF (`high flux') to AI (`intermediate') mode, 
the intensity loss is about a factor of $\sim{3}$. 
Going from AI to HR (`high resolution'), the loss factor is $\sim{4}$. 
These modes pair different slits in the double disk chopper.}
\label{reso}
\end{figure*}

During the early years of operation it became clear that achieving higher flux 
on sample -- at the expense of energy resolution -- was more important. 
Therefore, a new Fermi chopper rotor was designed and installed during a 
facility shutdown in 2014. 
Like the original rotor, it features two slit packages on the same vertical 
axis of rotation, which can be selected with a vertical translation of the 
chopper. 
The main difference between the old and the new slit packages is that the new 
slit packages are coarser than the old ones. 
As a result, the burst times of the two high-speed choppers are now better 
matched, in typical run conditions, to the respective distances of the 
choppers to the detector, 
improving the intensity vs energy resolution relationship.~\cite{Lec91}
The intensity gain (for the same resolution) is about $\sim{25}$\% on average. 
This gain outweighs the shift in the resolution range (towards coarser 
resolution) in which one can operate, because only 
very few experiments require the best available resolution. 
A current measurement of the energy resolution at the elastic line, using a 
vanadium reference sample, is shown in Fig.~\ref{reso}.


\subsection{Dedicated cryostat}

CNCS has a dedicated orange helium flow cryostat with a working temperature 
range of $\sim{1.8}-360$~K. 
The diameter of the variable temperature insert (VTI) is 100~mm. 
For studies with magnetic samples, the cryostat is often used with a $^3$He 
insert that extends the available temperature range on the lower end to 
$\sim{0.24}$~K. 
The insert routinely holds base temperature for more than 5 days without a 
need to recondense.
The VTI in the cryostat features a custom Cd liner that shields scattering 
directions without detector coverage. 
This is an essential feature to reduce background from incoherent 
backscattering or near Bragg-edge scattering in the aluminum of the cryostat 
tail. 
The Cd shield has a dedicated entry port for the beam and also a beam 
stop on the other side (inside the VTI). 
This implies that the cryostat can not be rotated on the sample table. 
Therefore single crystal samples are usually mounted on a stick that 
can be rotated inside the cryostat. 
Since data are recorded in event mode, this rotation can be continuous while 
counting scattered neutrons, as opposed to the more traditional way of 
recording individual runs sequentially at fixed positions of the sample 
rotation axis. 
Both these modes are operational and in use. 

\begin{figure*}[t]
\begin{center}
\includegraphics[width=6in]{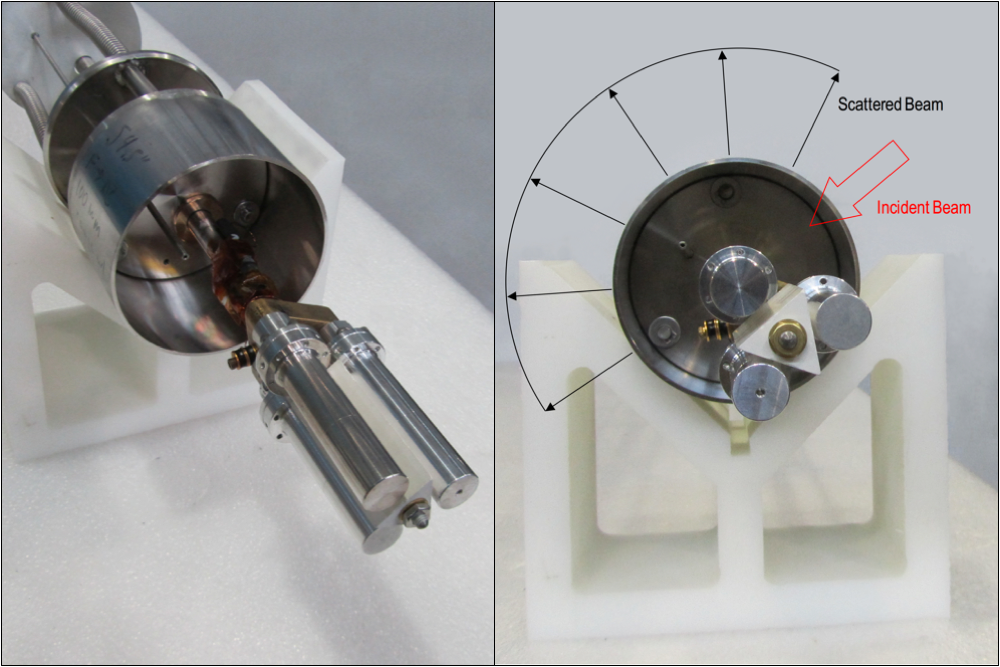}
\end{center}
\caption{The 3-sample changer stick photographed from below. 
The white triangular piece is made from absorbing boron nitride. 
The powder cans are 5/8'' in diameter, a standard size at CNCS that 
matches the beam width. 
These cans are typically used with annular inserts. }
\label{stick}
\end{figure*}

The large diameter of the cryostat's VTI also enabled the design of a 
3-sample changer stick that is shown in Fig.~\ref{stick}. 
The purpose of this device is to mount three powder samples simultaneously 
and to save time with temperature changes when several samples are being 
measured in an experiment. 
The three samples can be rotated around an off-center vertical axis 
while the cryostat is stationary. 
This motion allows to put one sample in the beam while the other two 
are on the side. 
The triangular piece between the three samples (see Fig.~\ref{stick}) is made 
of boron nitride which absorbs neutrons, thus shielding the two unused samples 
further from the beam and reducing the potential for secondary scattering. 
This device is very popular with users who measure many samples at several 
temperatures in one experiment. 
A future upgrade could add a second stage of three samples below the existing 
set of three, such that a total of six samples would be in the cryostat 
at the same time. 
One would likely have to use somewhat shorter powder cans than the current 
standard design. 
A vertical travel of the cryostat 
of the order of $\sim{5}$~cm during a measurement is feasible. 


\subsection{Planned future development -- polarized beam}

A first test experiment with polarized beam has been attempted at CNCS but more 
development work is needed before such an option can be offered to the user 
program. 
The setup made use of polarized $^3$He cells both for polarizing the beam and 
for analyzing the polarization of the scattered beam. 
On the analyzer side, there is virtually no choice because of the desired 
large solid angle. 
The incident beam can also be polarized with a transmission polarizer based 
on a supermirror, but this is much harder to do because more modifications 
of the existing infrastructure would be necessary. 

It has been recognized that an inelastic time-of-flight spectrometer with 
wide angular coverage in combination with 
polarized beam would be of great value for studies on quantum critical 
phenomena, topological states of matter, quantum magnets, unconventional 
superconductors, and geometrically frustrated magnets.~\cite{lbl} 
Efforts at the Institut Laue-Langevin (ILL),~\cite{Ste06} 
the ISIS facility at the STFC Rutherford Appleton Laboratory 
(United Kingdom),~\cite{Bee11} 
the Japan Proton Accelerator Research Complex (J-PARC)~\cite{Yok14} and 
SNS~\cite{Win15} are all aimed at increasing the area of solid angle covered 
by the detector with simultaneous polarization analysis.  
With the exception of D7 at the ILL~\cite{Rul07} (which has an option to run 
with a Fermi chopper) and HYSPEC at SNS~\cite{Win15}, inelastic neutron 
scattering with polarized beam is currently limited to triple-axis type 
instruments which cannot cover large areas of $(Q,\omega)$ space in adequate 
time. 
The ongoing trend in materials science towards more complexity, however, means 
that broad surveys in reciprocal space will be increasingly needed to identify 
the key dynamical signatures in the scattering. 
An unambiguous separation of lattice and spin dynamics will become a key 
requirement, and `$xyz$'-polarization analysis is the only available technique 
known to allow this at all scattering angles simultaneously.~\cite{xyz13} 


\section{Scientific use of the instrument}
\label{use}

Like most neutron sources, SNS runs an international user program and the 
available beam time is awarded to proposals after a competitive review process 
has been held. 
At CNCS this competition is particularly strong, and only $\sim{20}$\% of all 
requests for beam time can be accommodated. 
The use of the beam time is driven by community demand. 
The vast majority of experiments conducted at CNCS address topics in hard 
condensed matter, such as magnetism, correlated electrons, superconductivity, 
phonons and heat transport, and energy materials. 
Other science areas such as protein and polymer dynamics, glass transition 
(boson peak), or the dynamics of atoms and molecules in confined 
geometry combine for $\sim{15}$~\% of the time used.

Nearly $\sim{75}$~\% of the beam time at CNCS is used to measure collective 
excitations in single crystals. 
Single crystal measurements play to the full strength of the instrument 
with its large detector area, accessing all three spatial directions 
simultaneously, and the ability to adjust the measurement range and 
resolution to the need of the particular system studied.  
With the focusing guide the $Q$ resolution in the vertical direction is 
relaxed but still good enough to measure a dispersion in this direction.


\subsection{Measurements at pressure}
\label{pressure}

Pressure is a relevant thermodynamic variable for many materials. 
For example, applied pressure has a profound effect on the critical temperature 
in many superconductors,~\cite{Jen58,Gao94,Che10} and many materials are known 
to be superconductors only under pressure.~\cite{Dro15} 
This can be understood considering that the parameters of a material which are 
mainly important for its superconducting properties, namely the electronic 
density of states at the Fermi energy, the phonon frequency spectrum, and the 
electron-phonon coupling, all may be sensitively pressure-dependent. 
Metal-insulator transitions are another example for a transition of the 
electronic state in a material that can be induced by 
pressure.~\cite{Ma09,Nau15,Sho15}

Neutron scattering under pressure presents challenges, and inelastic 
scattering in particular must be considered very difficult. 
The main reason is that the available sample volume is inevitably small. 
Neutron scattering is an intensity limited technique, and while diffraction 
can be done with very small samples,~\cite{Neu12,Boe13} the volume limitation 
has in the past often pushed the feasibility limit beyond what is possible 
with inelastic scattering. 
Another difficulty lies in the relatively large amount of material that makes 
up the equipment that applies pressure, which cannot all be shielded from the 
beam and causes significant background scattering. 
For these reasons, one only finds a limited number of inelastic neutron 
scattering studies in the literature, mostly in cases when gas or liquid cells 
could be used which allow to apply modest pressure, $\lesssim{0.5}$~GPa, 
to a larger sample volume, and on materials that scatter 
well.~\cite{Bla75,Blo75,Eck76,Sch84,Str04} 
CuBe clamp cells, McWhan cells and sapphire cells have also been occasionally 
used successfully for inelastic scattering.~\cite{Moo82,Yam84,Iva95}
Scattering from magnetic systems face the additional challenge that the 
sample must be cooled, most often to cryogenic temperatures. 
Spin waves in elemental magnets Tb~\cite{Kaw94} and Fe~\cite{Klo00} have 
been studied at pressure but the involved temperatures were relatively high, 
90~K and 300~K, respectively, whereas in the pioneering work on 
single-crystal FeCl$_2$~\cite{Vet75} a base temperature of 4~K was reached. 
Crystal field excitations in magnets have also been occasionally studied at 
high pressure~\cite{Vet77,Mes90,Str03}.

At CNCS, gas pressure cells have been used up to $\sim{0.5}$~GPa. 
The cells are made from Al alloy and can be held in the cryostat at base 
temperature. 
The pressure in the gas cell can be applied in-situ. 
For example, the research on supercooled confined water performed recently at 
CNCS shows what can be achieved with theses cells. 
The dynamics of deeply cooled water in a nano-porous silica matrix (MCM-41) 
was studied in the temperature range $\sim{160}$~K to $\sim{230}$~K at 
pressures up to $\sim{0.5}$~GPa, using a gas pressure cell.~\cite{Zhe14,Zhe15} 
The results of these experiments were interpreted as to show the existence of 
two distinct liquid phases in this range of pressure and temperature, which 
differ in their densities and local structures of the water. 
Particularly it was shown  that the behavior of the Boson peak of the 
confined water is strongly correlated to the low-density and high-density 
phases of the confined water. 
These studies significantly deepen our understanding of the thermodynamic 
behavior of supercooled confined water.

\begin{figure*}[t]
\begin{center}
\includegraphics[width=5in]{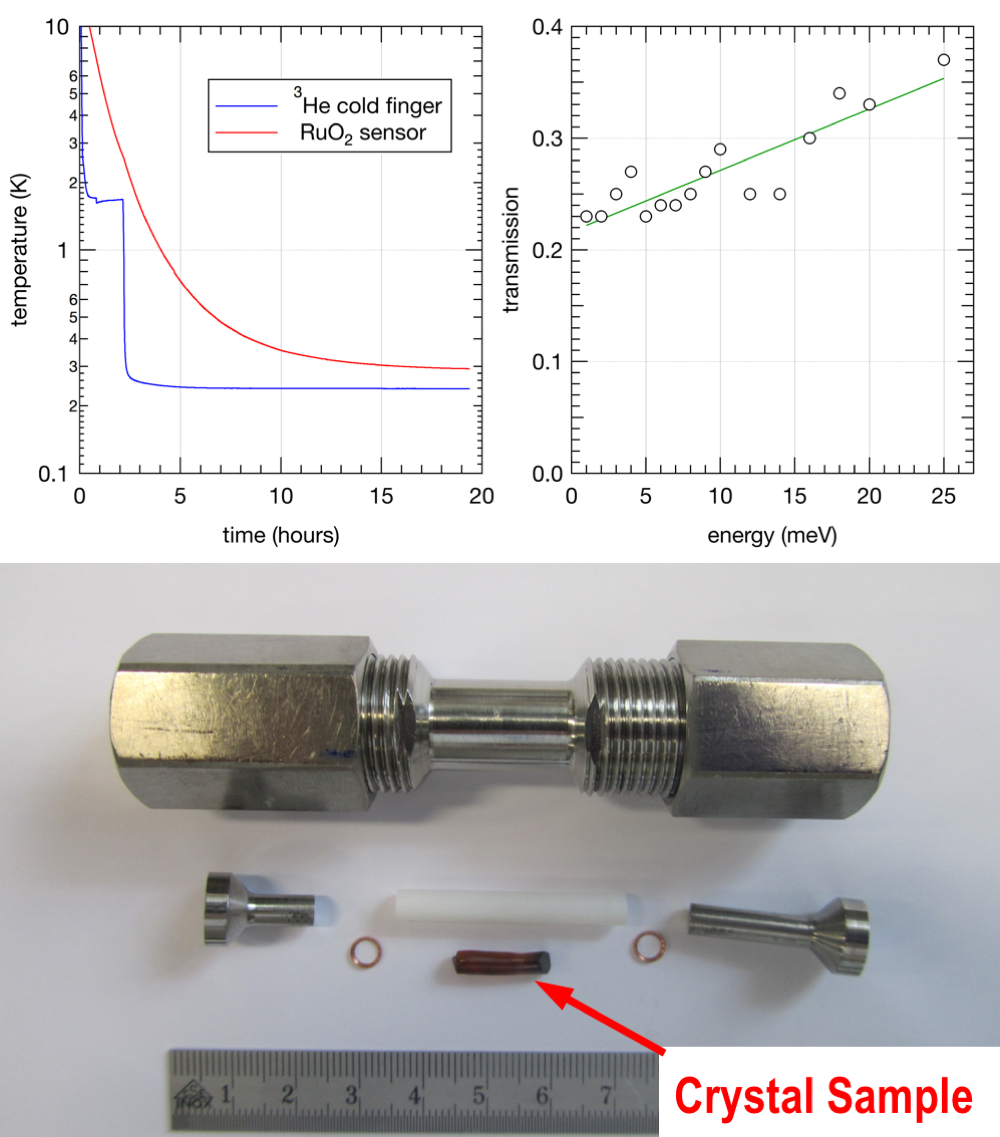}
\end{center}
\caption{The clamp pressure cell in use at CNCS. 
Upper part: Characterization of the thermal contact between sample 
and cold finger (left), and neutron transmission measured at CNCS (right). 
Lower part: Photograph of the cell and inner components. 
The crystal sample is about 4 mm in diameter and 15 mm long. 
This cell fits a standard dilution refrigeration insert which can be 
placed in the bore of a cryomagnet.}
\label{clamp}
\end{figure*}

In order to access higher pressure in a volume large enough for inelastic 
neutron scattering, modern clamp cells are now available that have been 
designed with the particular needs of scattering experiments in mind, 
in terms of materials choices, the amount of material in the beam, and 
the ability to cool the cell to cryogenic temperatures.~\cite{Rid14,Gut15}
A clamp cell currently in use at CNCS, which can reach pressure up to 
$\sim{2.0}$~GPa in a volume of $\sim{250}$~mm$^3$, is shown in the lower 
part of Fig.~\ref{clamp}. 
A test in a $^3$He insert has shown that the thermal contact to the sample 
(through teflon capsule and pressure medium, fluorinert) is sufficient to 
cool the sample down to $\sim{0.3}$~K. 
The test was conducted offline with a calibrated RuO$_2$ sensor in the sample 
position. 
The clamp cell has to be warmed and removed from the cryostat in order to 
change pressure. 
The total turnaround time between base temperature measurements at different 
pressures is $\sim{6}$~hours with the cryostat alone (base 1.8~K) and 
$\sim{12}$~hours with the $^3$He insert (base 0.3~K). 
The clamp cell is also small enough to fit the standard size of a dilution 
insert (diameter 32~mm) that can be placed in the bore of an 8~T cryomagnet 
available at SNS. 
One can thus measure a sample of $\sim{4}$~mm diameter, $\sim{20}$~mm long, 
at CNCS at 2~GPa, 8~T and 50~mK simultaneously. 
While it is technically possible to conduct such a measurement, a successful 
experiment under these conditions has yet to be performed. 
The pressure cell (walls of steel or hardened alloy) with fluorinert and 
teflon capsule, the magnet with its reduced outgoing divergence, and the 
dilution insert with another heat shield all add scattering material to the 
setup and reduce the signal to background ratio. 

Using a clamp cell of this design, the pressure dependent magnetic excitation 
spectrum in the $S=1/2$ quasi-two-dimensional gapped quantum antiferromagnet 
(C$_4$H$_{12}$N$_2$)Cu$_2$Cl$_6$ (PHCC) was studied at CNCS.~\cite{GPe15} 
The single crystal used in the CNCS experiment weighed ~150 mg. 
The pressure cell was mounted inside a helium flow cryostat and data were 
collected with incident energy of $E\nsub{i}=4.2$~meV at $T=1.5$~K. 
The full $S(Q,\omega)$ map was recorded by making 180~deg. rotations with 
1~deg. step size. 
At ambient pressure and cryogenic temperature, PHCC does not order 
magnetically, and the ground state spin singlet is separated by a gap of 
$\sim{0.98}$~meV from an $S=1$ triplet. 
Applying pressure, the gap can be reduced, and it was shown that at 
0.9~GPa and above, the excitations are gapless. 
This agreed with the findings of an independent $\mu^{+}$SR experiment 
which observed a quantum critical point at $p=0.43$~GPa and long-range 
magnetic order above. 
Thus it could be shown that the quantum phase transition in PHCC under 
pressure is driven by the weakening of a single Cu-Cu superexchange pathway. 
Besides, at high pressure a sizeable spin wave dispersion along the 
interlayer direction was observed implying substantial three-dimensional 
correlations.

\begin{figure*}[t]
\begin{center}
\includegraphics[width=6in]{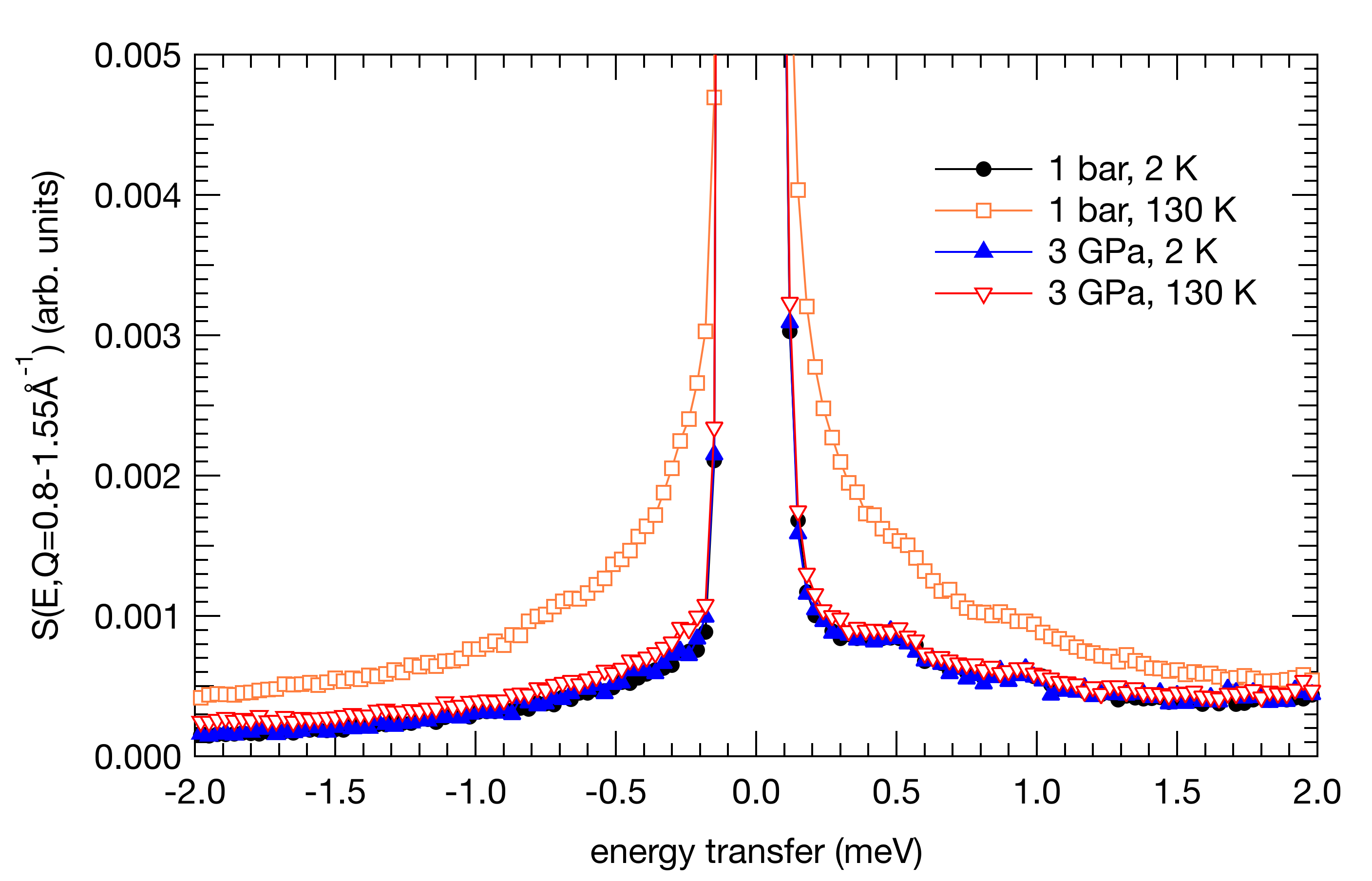}
\end{center}
\caption{Inelastic neutron scattering spectra of water in hemimorphite 
measured at CNCS at different temperatures and pressures.  
Strong QENS broadening is visible for hemimorphite at 130~K and ambient 
pressure due to the fast rotational diffusion of water molecules, 
while the data for $p=3$~GPa almost coincide for $T=2$~K and $T=130$~K, 
showing the absence of QENS broadening through the entire temperature range. }
\label{hemi}
\end{figure*}

A larger clamp cell, that pushes the achievable pressure to $\sim{3}$~GPa, 
was used to investigate the pressure dependence of fast rotational 
diffusion of water molecules in the mineral hemimorphite 
Zn$_4$Si$_2$O$_7$(OH)$_{2}\cdot$H$_2$O at cryogenic temperatures.  
At ambient pressure the mineral belongs to the orthorhombic space 
group $Imm2$.~\cite{Hil77} 
The structure consists of rings of corner-sharing ZnO$_4$ and SiO$_4$ 
tetrahedra which make up channels.  
The water molecule occupies, on average, a symmetrical position in the 
channels, resting entirely upon the crystallographic $a-c$ plane. 
A water molecule forms four (almost planar) hydrogen bonds with hydroxyl 
groups in the hemimorphite framework. 
Application of hydrostatic pressure above 2.5~GPa causes the mineral to 
undergo a structural phase transition from $Imm2$ to $Pnn2$.~\cite{Ser11} 
The asymmetric deformation of the channels leads to a shorter H$_2$O-OH 
contact, and the planar hydrogen bond network becomes tetrahedrally 
coordinated.  
By changing the local crystal environment, the application of high 
pressure to hemimorphite is thus expected to profoundly alter the 
diffusive dynamics of confined water molecules. 
Fig.~\ref{hemi} shows quasi-elastic neutron scattering (QENS) spectra of 
hemimorphite under hydrostatic pressure measured at CNCS at 2~K and 
130~K (with an incident neutron energy $E\nsub{i}=3$~meV). 
At low temperatures, water dynamics in hemimorphite is undetectable 
at ambient pressure and at $p=3$~GPa, and the measured $S(E)$ represents 
the resolution function of the spectrometer.  
At $T=130$~K the spectrum of hemimorphite at ambient pressure shows 
strong QENS broadening due to fast rotational diffusion of water, 
while the data for $p=3$~GPa are very similar to those at 2~K. 
These measurements demonstrate that water in hemimorphite in the high 
pressure phase (being connected to the hemimorphite cage via tetrahedrally 
coordinated hydrogen bonds) does not show fast rotational diffusion 
like at ambient pressure where the water being coordinated via 
planar hydrogen bonds.


\subsection{User experiments}
\label{user}

One of the greatest strengths of CNCS is the ability to measure collective 
excitations in crystals, simultaneously, in the energy domain and in all three 
spatial directions, with adjustable energy- and $Q$-resolution. 
One of the science areas that exploits this ability is the research on 
thermoelectric materials. 
Here, the main question addressed with inelastic neutron scattering is that 
of the microscopic origin of the unusual strong scattering of the heat 
carrying phonons in materials such as PbTe~\cite{Del11,CLi14}, 
AgSbTe$_2$~\cite{Ma13,Ma14}, SnTe and SnSe~\cite{Li14,Li15}. 
The dispersion and line widths of the phonons need to be measured accurately, 
in various directions, and with good $Q$ resolution. 
It is also essential to complement the measurements with density functional 
theory (DFT) calculations in order to achieve a consistent understanding of 
the measurements. 
The experiments found that the origin of the strong phonon scattering is 
quite different in these materials. 
In PbTe a strong anharmonic repulsion was observed between the ferroelectric 
transverse optic phonon and the longitudinal acoustic modes. 
This is the signature of an underlying anharmonic interaction between these 
phonons.
In AgSbTe$_2$ it is an atypical degree of complexity of the crystal 
structure at the microscopic level that was identified as the source of an 
unusual level of phonon scattering. 
This material forms nano-sized domains which differ in the near neighbor 
ordering of some of the ions. 
The phonons cannot propagate undisturbed through this structure, which 
manifests itself in a much broadened phonon line width when compared to 
PbTe for example. 
SnSe on the other hand is near a lattice instability, leading to strongly 
anharmonic interaction potentials between Sn and Se.

Excitations in quantum magnets with low energy scales are another research 
area in which CNCS is designed to make 
contributions.~\cite{Wan12,Sch13,Mat14,Wu16}
For example, geometrically frustrated magnets~\cite{Cla14,Mac15,Ma16} are 
often characterized by a macroscopically degenerate ground state, and for 
this reason tend to have a high number of fluctuation modes at low energy 
that may persist to very low temperature. 
The good energy resolution provided by cold neutrons is therefore ideally 
suited to studying the spin fluctuations in such systems. 
In multiferroic materials~\cite{Ye11,Fro11,Ehl13} the energy 
scales are also often low.

A magnetic field is another external parameter with which a spin system may 
be manipulated. 
The critical magnetic fields associated with low energy phenomena are also 
often low and technically within reach, which is a good match for an 
instrument like CNCS. 
Fields up to 16~T have been reached at CNCS in the past.~\cite{Fri15} 
However, integrating a split pair magnet into an instrument such as CNCS 
also does present challenges. 
Split pair magnets are the most commonly used design in neutron scattering 
applications, because the scattered beam can exit at any angle in the 
scattering plane. 
The design requirements of such magnets imply that the solid angle in which 
scattered neutrons can be detected must be restricted to some 
extent.~\cite{Bro09}
Modern design solutions allow one to trade off vertical divergence for 
scattering angle range and closely match the host instrument geometry, 
thus minimizing intensity losses. 
The 8~T magnet currently in use at the SNS facility allows for an outgoing 
divergence of the scattered beam of $\pm{12}^{\circ}$ which is a significant 
improvement over previous designs (the scattered beam can be detected in a 
solid angle of nearly 2~sr).  
It also avoids the use of aluminium spacers in the scattered beam.  
The inner bore has a diameter of 34~mm at the sample location and the beam 
can be about 30~mm tall at this point. 
A recent development that has the potential to extend the accessible field 
range, when combined with time-of-flight methods, is that of pulsed 
magnets.~\cite{Noj11}

Both magnetic and structural excitations are of interest in studies of the 
unconventional supercondutivity seen in materials with a structural motif of 
FeAs, FeSe or 
BiS$_2$ layers.~\cite{Cru08,Chr08,Lum10,Miz12,Kim13,Lee13,Lee14,Li16} 
Many of the recently discovered `unconventional' superconductors possess 
magnetic ions and long range magnetic order in at least one thermodynamic 
phase. 
Superconductivity is typically induced by doping charge carriers, which at 
the same time reduces the magnetic ordering temperature or suppresses the 
ordering completely. 
A common feature is the existence of a resonant magnetic excitation within 
the superconducting phase, which is localized in both energy and wavevector. 
Such excitations can be directly seen with inelastic neutron scattering, 
but it is to date not fully understood how exactly they are connected to 
superconductivity in these materials.
This is a field of strong current interest in condensed matter physics. 


While the majority of experiments conducted at CNCS are hard condensed matter 
studies, quasielastic scattering studies in soft matter materials make up a 
non-negligible part of the science program at CNCS. 
In the area of polymer and protein dynamics, various studies focused on 
the role of hydration water,~\cite{Nic12,Nic16} the nature of the collective 
vibrations (``boson peak''),~\cite{Nic13,Per14} and the role of the secondary 
structure for the rigidity and functionality of these 
molecules~\cite{Per13,SPe14,Nic15}. 
These experiments are generally {\em not} intensity limited, 
unless the samples are fully deuterated. 
A full spectrum may be collected at CNCS in less than an hour on a 100~mg 
sample. 
An upgrade offering polarized beam would therefore be particularly beneficial 
for this science, as it would enable one to disentangle coherent (collective, 
many-particle) and incoherent (single-particle) dynamical modes in soft 
(hydrogenous) materials, which may overlap in the time 
domain.~\cite{Far02,Bur14} 
This would make it possible to rigorously test model assumptions that one 
currently has to make in order to separate the different processes in 
measurements with unpolarized neutron beams.

It is known that the dynamical behavior of atoms and molecules in confined 
geometry can be substantially different from the bulk.~\cite{Ber00} 
Rather than near correlation shells of its own species, an individual 
molecule will see a material-dependent wall potential.
For example, the dynamics of H$_2$O confined in $\sim{5}$~{\AA} diameter 
channels of beryl single crystal have been studied by using QENS and inelastic 
neutron scattering (INS) at CNCS and SEQUOIA.~\cite{Kol14,Kol16}
The QENS study with energy resolution 0.25~meV with the scattering momentum 
transfer along the channels showed gradual freezing of water molecule 
dynamics at temperatures below $\sim{200}$~K, whereas the dynamical features 
was generally much weaker with the momentum transfer perpendicular to the 
channels.
At higher temperatures the data were described as two-fold rotational jumps 
about the axis coinciding with the direction of the dipole moment 
(perpendicular to the channels), with a residence time of 5.5~ps at 225~K. 
The INS spectra of water in beryl measured at CNCS and SEQUOIA in the 
direction perpendicular to the channels revealed a number of peaks which 
were uniquely assigned to water quantum tunneling. 
In addition, the water proton momentum distribution measured with deep 
inelastic neutron scattering (VESUVIO, ISIS~\cite{And05}) directly showed 
coherent delocalization of the water protons in the ground state. 
The observation of multiple tunneling peaks and a coherent delocalization 
of protons over all possible positions across the beryl channel allowed the 
view of the tunneling water as ``a new state of the water 
molecule''.~\cite{Kol16}


\section{Conclusion}
\label{Conclusion}

CNCS offers excellent beam intensity and resolution for inelastic and 
quasi-elastic neutron scattering experiments in the cold and thermal 
neutron energy ranges. 
The instrument performance allows to push the feasibility limits for 
successful user experiments towards smaller samples ($\sim{100}$~mg of 
non-hydrogenous material) and high pressure (in combination with strong 
magnetic field and/or very low temperature, if desired). 
The instrument has now a solid track record of answering questions at the 
forefront of many areas in condensed matter physics, such as 
quantum magnets, 
unconventional superconductors, 
geometrically frustrated magnets,
thermoelectric materials, 
polymer and protein dynamics,
and matter in confined geometry.

\section{Acknowledgements}
\label{Acknowledgements}

Research at ORNL's Spallation Neutron Source is sponsored by the 
Scientific User Facilities Division, Office of Basic Energy Sciences, 
U.S. Department of Energy. 
The authors would like to thank E. Iverson for help with the beam image, 
and the technical staff at SNS, S. Elorfi, C. Fletcher, C. Redmon, for 
their support with sample environment equipment operation.
Special thanks are due to H. Ambaye, E. Brophy, J. Carruth, M. Everett, 
L. Jones, J. Niedziela, 
and A. Parizzi for their help with instrument operation. 

\section{References}
\label{References}

\bibliographystyle{unsrt} 
\bibliography{CNCS}




\pagebreak



\pagebreak


\pagebreak


\pagebreak


\pagebreak


\pagebreak


\end{document}